\documentclass[11pt]{article}
\usepackage{moriond,epsfig}

\def\Journal#1#2#3#4{{#1} {\bf #2}, #3 (#4)}

\def\NIMA{{\em Nucl. Instrum. Methods} A}

\def\PLB{{\em Phys. Lett.}  B}
\def\PRL{\em Phys. Rev. Lett.}
\def\PRD{{\em Phys. Rev.} D}

\def\EPJ{\em Eur. Phys. J. direct}

\def\be{\begin{equation}}
\def\ee{\end{equation}}
\def\bea{\begin{eqnarray}}
\def\eea{\end{eqnarray}}

\begin{document}
\vspace*{4cm}
\title{PENGUIN-MEDIATED $B$ DECAYS AT BELLE}

\author{AKIMASA ISHIKAWA\\(for the Belle Collaboration)}

\address{Department of Physics, Nagoya University, Furo-Cho, Nagoya, Japan}

\maketitle\abstracts{
We report on the results of penguin-mediated $B$ decays at the Belle experiment. The analyses were based on approximately 32 million $B\overline{B}$ events collected at the $\Upsilon(4S)$ resonance with the Belle detector at the KEKB $e^+e^-$ storage ring. The $b \to s \gamma$ transition was studied through exclusive decays: $B \to K^{*} \gamma$, $B^0 \to K_{2}^{*}(1430)^0 \gamma$, $B^+ \to K^+\pi^-\pi^- \gamma$, $B^+ \to K^{*0} \pi^+ \gamma$ and $B^+ \to K^+\rho^0 \gamma$. 
The $b \to s \ell^+ \ell^-$ transition was searched through both exclusive decays, $B \to K^{(*)} \ell^+ \ell^-$, and inclusive decay, $B \to X_{s} \ell^+ \ell^-$. 
We observed the decay processes $B^+ \to K^+\pi^+\pi^- \gamma$ and $B \to K \ell^+ \ell^-$ for the first time.
}
\section{Introduction}
In the Standard Model (SM), flavor-changing neutral current (FCNC) decays are forbidden at tree level. However, FCNC decays are induced through loop diagrams, such as penguin diagrams or box diagrams. These loop diagrams are sensitive to new physics, since heavy particles beyond the SM, such as charged Higgs or SUSY particles, contribute to additional loop diagrams,  branching fraction or kinematic variables can be deviated from the SM values. 

We have studied the FCNC decays of $b \to s \gamma$ and $b \to s \ell^+ \ell^-$ using a data sample collected with the Belle detector~\cite{Belle} at the KEKB~\cite{KEKB} storage ring. The data sample corresponds to 29~fb${}^{-1}$ taken at the $\Upsilon(4S)$ resonance and contains approximately 32 million $B\overline{B}$ pairs.

\section{$b \to s \gamma$ transition}

\subsection{Analysis of $B \to K^{*} \gamma$}
\begin{figure}[bpht]
\begin{center}
\psfig{figure=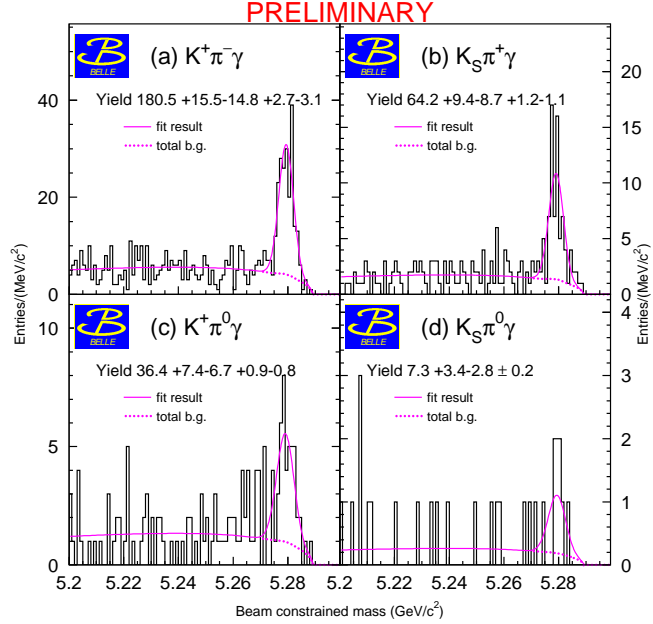,height=9cm}
\end{center}
\caption{$M_{\rm{bc}}$ distribution of $B \to K^*(892) \gamma$.\label{fig:kstgamma}}
\end{figure}

Exclusive $B \to K^{*} \gamma$ 
decays were reconstructed from a high energy photon and a $K^*$ ($K^{*0} \to K^+\pi^-$ or $K_S\pi^0$, $K^{*+} \to K_S\pi^+$ or $K^+\pi^0$).
The continuum background was suppressed by a likelihood ratio, which was constructed from an event shape variable Super Fox-Wolfram (SFW) and the $B$ meson flight direction. After applying a likelihood ratio cut, the $K^{*} \gamma$ final state was cleanly reconstructed. The beam energy constrained mass ($M_{\rm{bc}}$) distributions for each sub decay modes are shown in Fig~\ref{fig:kstgamma}. We obtained the branching fractions:
\begin{eqnarray}
	{\cal B}(B^{0} \to K^{*0} \gamma) &=& ( 4.08^{+0.35}_{-0.33}\pm0.26 ) \times 10^{-5},\\
        {\cal B}(B^{+} \to K^{*+} \gamma) &=& ( 4.92^{+0.59}_{-0.54}{}^{+0.38}_{-0.37} ) \times 10^{-5}.
\end{eqnarray}

We also checked for any partial decay rate asymmetry. We only used the self-tagging modes: $K^{\pm}\pi^{\mp}\gamma$, $K^{\pm}\pi^0\gamma$ and $K_{S}\pi^{\pm}\gamma$. The wrong tag fraction due to hadron misidentification was estimated to be only 1.2\%. We determined the partial rate asymmetry as
\begin{eqnarray}
        {\cal A}_{CP} = \frac{\Gamma(\overline{B} \to \overline{K}^{*} \gamma)-\Gamma(B \to K^{*} \gamma)}
			{\Gamma(\overline{B} \to \overline{K}^{*} \gamma)+\Gamma(B \to K^{*} \gamma)} = ( +3.2^{+6.9}_{-6.8}\pm{2.0} )\%,
\end{eqnarray}
which corresponds to
\begin{eqnarray}
	(-8.5 < {\cal A}_{CP} < 14.9) \% \ \ \ \ ( \rm{90\%~C.L.} ),
\end{eqnarray}
and was found to be consistent with zero. This limit is the most stringent over previously published results~[\ref{ref:CLEOAcpK2stgamma},\ref{ref:BabarAcp}].

\subsection{Analysis of $B^0 \to K_2^{*}(1430)^0 \gamma$}
\begin{figure}[thp]
\begin{center}
\psfig{figure=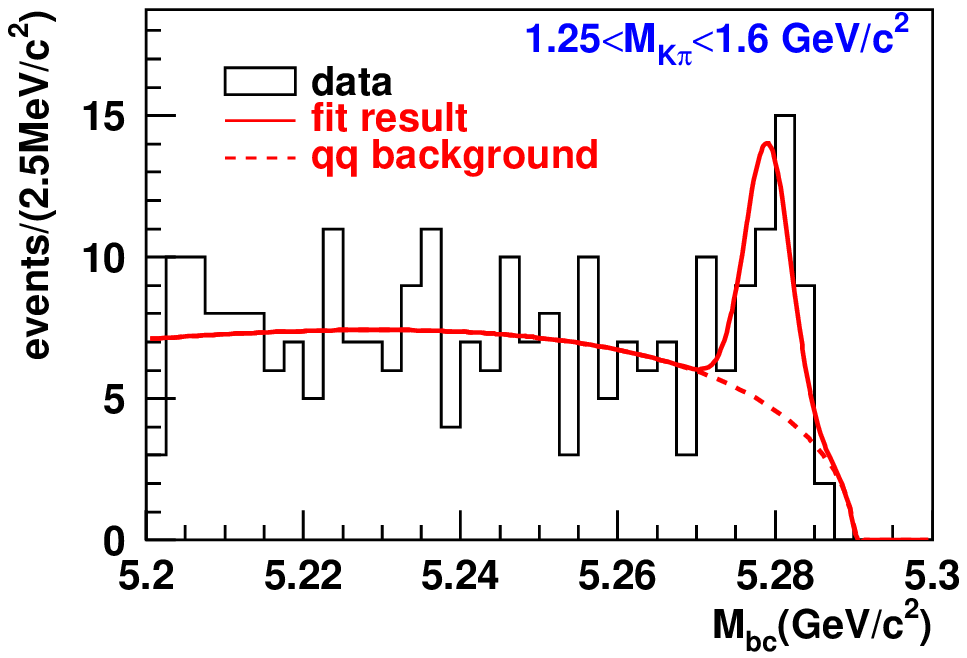,height=5cm}
\psfig{figure=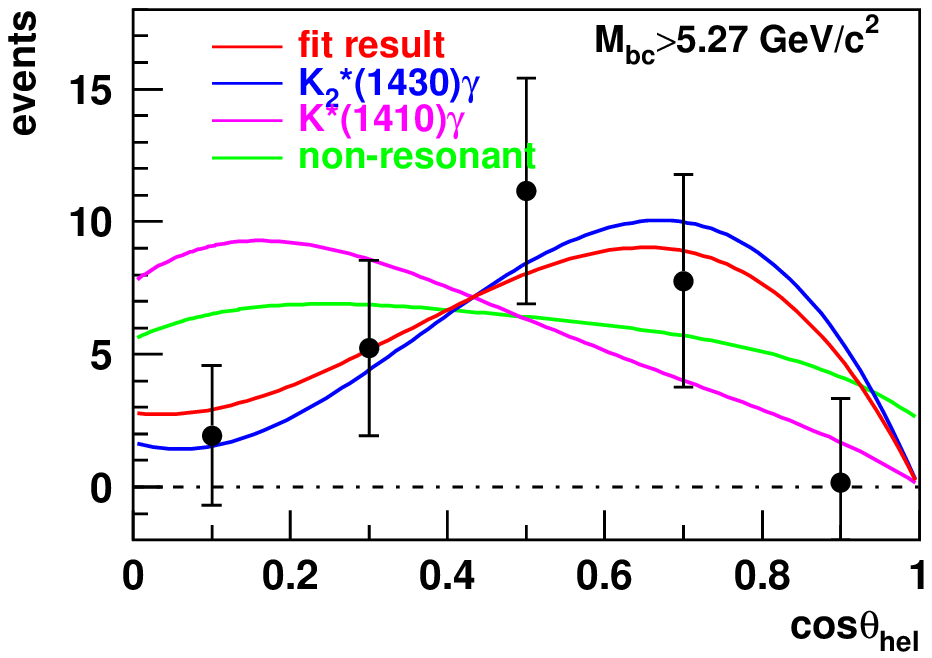,height=5cm}
\end{center}
\caption{$M_{\rm{bc}}$ distribution (left) and $\cos\theta_{\rm{hel}}$ distribution (right) in $B \to K^+\pi^-\gamma$ decay.\label{fig:mbck2stgammacoshel}}
\end{figure}

In the $B^0 \to K_2^{*}(1430)^0 \gamma$ analysis, the selection criteria used were similar to those used in the $K^* \gamma$ analysis. The $K_2^*(1430)^0$ was reconstructed from $K^+\pi^-$, which lies between  $1.25$ and $1.60$~GeV/$c^2$. We obtained $27.0\pm{6.7}^{+0.8}_{-3.4}$ events for the $B \to K^+\pi^- \gamma$ decay from the $M_{\rm{bc}}$ fit (Fig.~\ref{fig:mbck2stgammacoshel}). 

To distinguish the signal $B^0 \to K_2^*(1430)^0 \gamma$ from $K_1(1430)\gamma$ and non-resonant $K^+\pi^-\gamma$, we performed a unbinned maximum likelihood fit to $M_{\rm{bc}}$,the  helicity angle ($\cos\theta_{\rm{hel}}$) and the $K^+\pi^-$ invariant mass ($M_{K\pi}$). Fig.~\ref{fig:mbck2stgammacoshel} shows the $\cos\theta_{\rm{hel}}$ distribution, where the continuum background has been subtracted. We found that $K_2^*(1430)$ resonance was the dominant component and obtained a signal yield of $24.0^{+9.2}_{-8.5}{}^{+0.7}_{-1.4}$ events. The branching fraction of $B^0 \to K_2^{*}(1430)^0 \gamma$ was determined to be
\begin{eqnarray}
	{\cal B}(B^{0} \to K_2^*(1430)^0 \gamma) = ( 1.50^{+0.58}_{-0.53}{}^{+0.11}_{-0.13} )\times 10^{-5}.
\end{eqnarray}
This result is consistent with a previous measurement~\cite{CLEOAcpK2stgamma} and theoritical predictions.~\cite{K2stgamma}

\subsection{Analysis of $B^+ \to K^+ \pi^+ \pi^- \gamma$}
\begin{figure}[bhpt]
\begin{center}
\psfig{figure=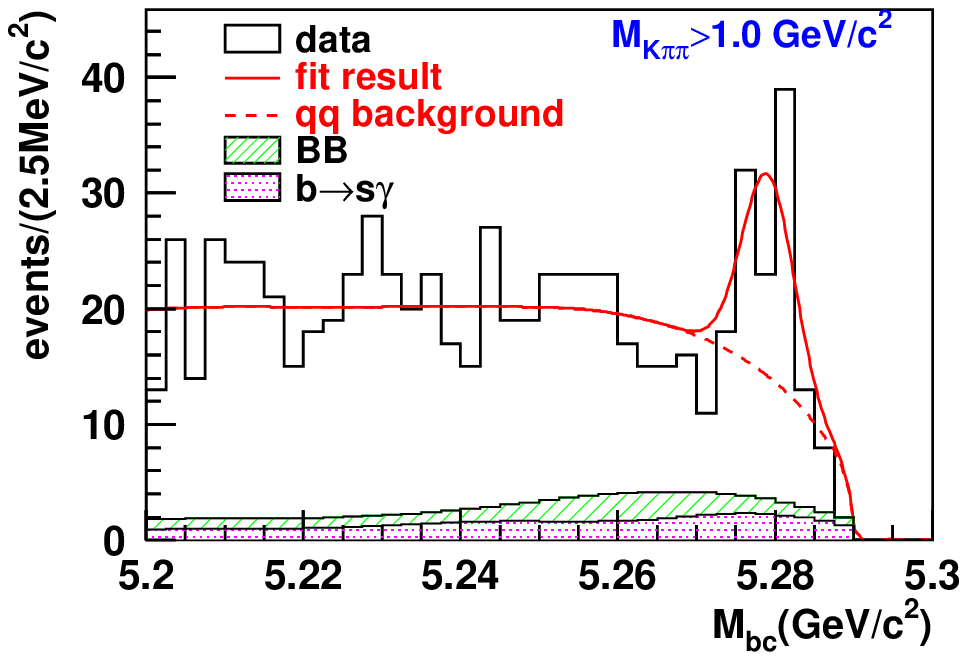,height=5cm}
\psfig{figure=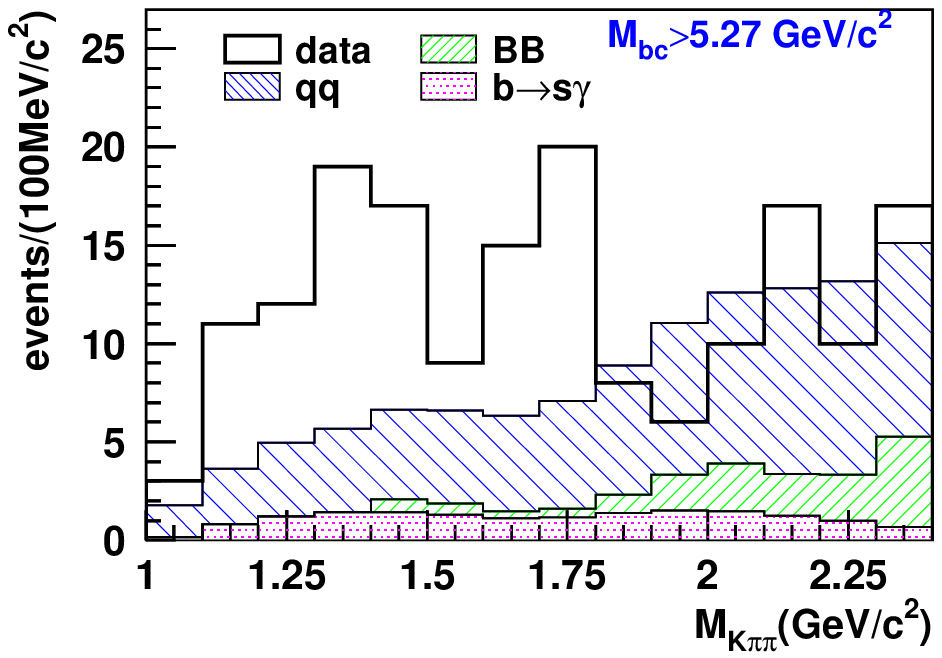,height=5cm}
\end{center}
\caption{The $M_{\rm{bc}}$ distribution of $B \to K\pi\pi \gamma$.\label{fig:mbckpipigamma}}
\end{figure}

\begin{figure}[ph]
\begin{center}
\psfig{figure=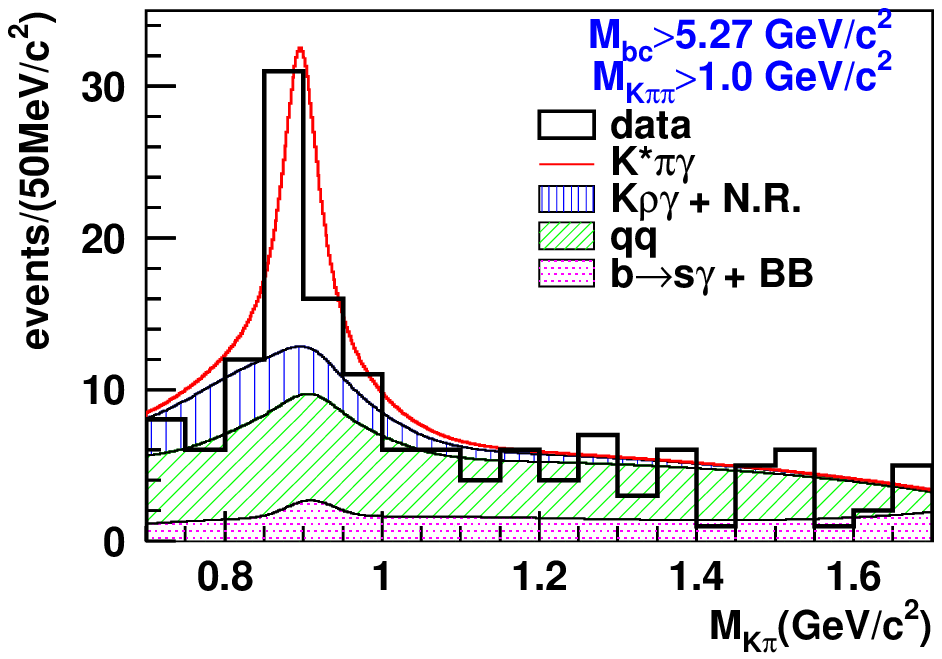,height=5cm}
\psfig{figure=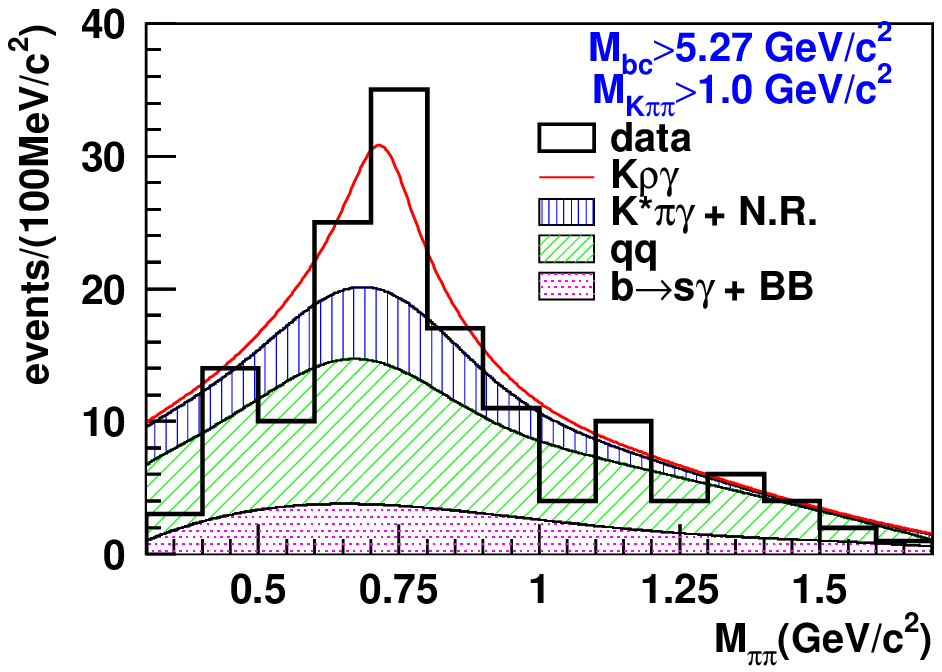,height=5cm}
\end{center}
\caption{The $M_{K\pi}$ (left) and $M_{\pi\pi}$ (right) distributions in  $B \to K\pi\pi \gamma$ decay.\label{fig:mkpimpipi}}
\end{figure}

We extended the same analysis to the three body hadronic final states of radiative decay. The $B^+$ was reconstructed from $K^+ \pi^+ \pi^- \gamma$. The invariant mass of $K^+ \pi^+ \pi^-$ was required to be between 1.0~GeV/$c^2$ and 2.4~GeV/$c^2$. To extract the signal yield, we made a fit to the $M_{\rm{bc}}$ distribution and obtained $57.7^{+11.8}_{-11.1}{}^{+6.4}_{-1.9}$ signal events with a statistical significance of $6.0 \sigma$ (Fig.~\ref{fig:mbckpipigamma}). We first observed $B^+ \to K^+ \pi^+ \pi^- \gamma$ and determined the branching fraction as
\begin{eqnarray}
	{\cal B}( B^+ \to K^+ \pi^+ \pi^- \gamma ) = ( 2.43^{+0.50}_{-0.47}{}^{+0.35}_{-0.23} )\times 10^{-5}.
\end{eqnarray}
The invariant mass of reconstructed $K^+ \pi^+ \pi^-$ is shown in Fig.~\ref{fig:mbckpipigamma}. We can observe a significant excess below $1.8$~GeV/$c^2$. In this mass region, there are many higher kaonic resonances which contribute to this decay mode, which are difficult to distinguish. However we can still identify $K^+ \pi^+ \pi^-$ via $K^* \pi$ or $K \rho$. We performed on unbinned maximum likelihood fit to $M_{\rm{bc}}$, the invariant mass of $K^+\pi^-$ ($M_{K\pi}$) and $\pi^+\pi^-$ ($M_{\pi\pi}$). Fig.~\ref{fig:mkpimpipi} shows the $M_{K\pi}$ and $M_{\pi\pi}$ distributions along with the fit. We obtained the signal yield of $K^* \pi \gamma$, $K \rho \gamma$ and non-resonant $K\pi\pi \gamma$ to be $32.5^{+10.8}_{-10.0}{}^{+1.9}_{-1.7}$, $24.2\pm11.6{}^{+3.4}_{-7.0}$ and $0.0^{+11.0}_{-0.0}$, respectively. The branching fraction and upper limits were found to be
\begin{eqnarray}
	{\cal B}( B^+ \to K^*0 \pi^+ \gamma ) = 2.04^{+0.67}_{-0.62}\pm0.22 ) \times 10^{-5},\\
	{\cal B}( B^+ \to K^+ \rho^0 \gamma ) < 1.9 \times 10^{-5} \ \ \ \ ( \rm{90\%~C.L.} ),\\
	{\cal B}( B^+ \to K^+ \pi^+ \pi^-  \gamma \ \ {\rm{non-resonant}} ) < 0.92 \times 10^{-5} \ \ \ \ ( \rm{90\%~C.L.} ).
\end{eqnarray}
This result shows that the $B^+ \to K^+ \pi^+ \pi^- \gamma$ process is consistent with a mixture of $B^+ \to K^*0 \pi^+ \gamma$ and $B^+ \to K^+ \rho^0 \gamma$.

We measured the exclusive two and three body hadronic final states of radiative decay. These exclusive decay rates could be compared with the inclusive $B \to X_s \gamma$ decay rate.~\cite{Xsgamma} When we calcurate the exclusive decay rate, the isospin invariance in the decay rate was assumed. We summed up those decay rates, giving $11.7\pm2.0 \times 10^{-5}$. This accounts for $35\pm11\%$ of the inclusive decay rate.

\section{$b \to s \ell^+ \ell^-$ transition}
\subsection{Analysis of $B \to K^{(*)} \ell^+ \ell^-$}
Here, we summarize what is already published in Ref.~\cite{Kll} We reconstructed $B \to K^{(*)} \ell^+ \ell^-$ from oppositely charged lepton pair ($e^+e^-$ or $\mu^+\mu^-$) and a kaon or a $K^*$ ($K^+$, $K_S \to \pi^+\pi^-$, $K^{*0} \to K^+\pi^-$ or $K_S\pi^0$, $K^{*+} \to K_S\pi^+$ or $K^+\pi^0$). The $J/\psi X_{s}$ and $\psi^{'} X_{s}$ events were vetoed by di-lepton invariant mass cuts. The di-electron which comes from photon conversion and $\pi^0$ Dalitz decay are removed by requiring $M_{ee} > 0.14$~GeV/$c^2$. The continuum background was suppressed using a likelihood ratio formed by a Fisher discriminant, $\cos\theta_B$ and the angle between the $B$ candidate sphericity axis and z axis $\cos\theta_{\rm{sph}}$. The Fisher discriminant was calculated from the energy flow in 9 cones along the $B$ candidate sphericity axis and the normalized second Fox-Wolfram moment ($R_2$). Another major background is $B\overline{B}$ events where both $B$ mesons decay to $X_c\ell\nu$. This background is suppressed by another likelihood ratio constructed from a missing energy in the event and $\cos\theta_B$.

\begin{figure}[ph]
\begin{center}
\psfig{figure=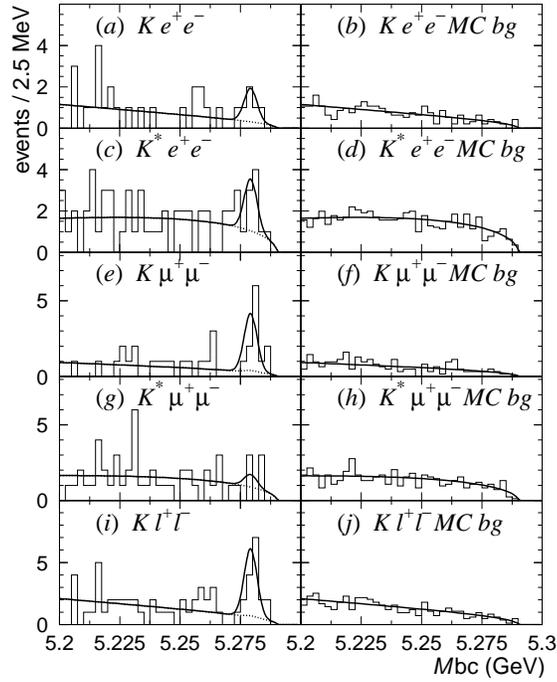,height=9cm}
\end{center}
\caption{$M_{\rm{bc}}$ distributions of $B \to K^{(*)} \ell^+ \ell^-$ decay. The left column is for the data and the right column is for the MC background.\label{fig:mbcfitbgshape}}
\end{figure}

\begin{figure}[ph]
\begin{center}
\psfig{figure=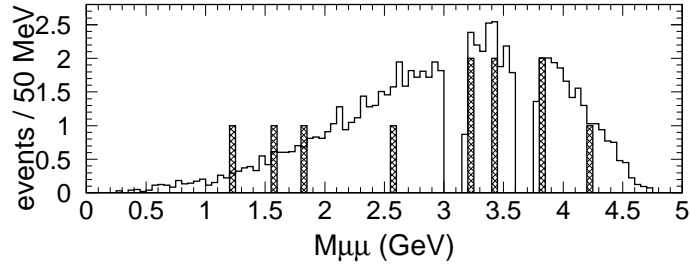,height=3.5cm}
\end{center}
\caption{$M_{\mu\mu}$ distribution of $B \to K^* \ell^+ \ell^-$. The hatched histogram shows the data distribution, while the open histogram shows the MC signal distribution. \label{fig:mll}}
\end{figure}

To extract the signal yield, we made a fit to the $M_{\rm{bc}}$ distributions (Fig.~\ref{fig:mbcfitbgshape}). We observed $9.5^{+3.8}_{-3.1}{}^{+0.8}_{-1.0}$ events for $B \to K \mu^+ \mu^-$ with a statistical significance of 4.7. The di-muon invariant mass distribution is shown in Fig.~\ref{fig:mll}, and is found to be consistent with the Monte Carlo (MC) expectation. For other modes, we observed no significant excess: $4.1^{+2.7}_{-2.1}{}^{+0.6}_{-0.8}$ events for $B \to K e^+ e^-$, $6.3^{+3.7}_{-3.0}{}^{+1.0}_{-1.1}$ events for $B \to K^* e^+ e^-$ and $2.1^{+2.9}_{-2.1}{}^{+0.9}_{-1.0}$ events for $B \to K^* \mu^+ \mu^-$. In the combined $K \ell^+\ell^-$ mode, we observed $13.6^{+4.5}_{-3.8}{}^{+0.9}_{-1.1}$ events with a statistical significance of 5.3. For the mode with significant signal events, we determined the branching ratios:
\begin{eqnarray}
	{\cal B}( B \to K \ell^+ \ell^- ) &=& ( 0.75^{+0.25}_{-0.21}\pm0.09 ) \times 10^{-6},\\
	{\cal B}( B \to K \mu^+ \mu^- )   &=& ( 0.99^{+0.40}_{-0.32}{}^{+0.13}_{-0.14} ) \times 10^{-6},
\end{eqnarray}
and found them to be consistent with the SM predictions.~\cite{Exclusive} For other modes, we set the upper limits as follows:
\begin{eqnarray}
	{\cal B}( B \to K e^+ e^- )       &<& 1.3 \times 10^{-6} \ \ \ \ ( \rm{90\%~C.L.} ),\\
	{\cal B}( B \to K^* e^+ e^- )     &<& 5.6 \times 10^{-6} \ \ \ \ ( \rm{90\%~C.L.} ),\\
	{\cal B}( B \to K^* \mu^+ \mu^- ) &<& 3.1 \times 10^{-6} \ \ \ \ ( \rm{90\%~C.L.} ).
\end{eqnarray}

\subsection{Analysis of $B \to X_{s} \ell^+ \ell^-$}
The $B \to X_{s} \ell^+ \ell^-$ decay was reconstructed by combining an $X_s$ system and an oppositely charged lepton pair ( $e^+e^-$ or $\mu^+\mu^-$ ). The $X_s$ is formed from one charged or neutral kaon and zero to four pions where at most one neutral pion is allowed. The continuum background is suppressed by $R_2$ and the he remaining tracks. The $B\overline{B}$ background is suppressed by two likelihood ratios. The first likelihood ratio is formed from the missing energy and the $X_s$ mass. The second likelihood is constructed from $\cos\theta_B$ and the sum of the cosine of the angle between the kaon and leptons ($\cos\theta_{K\ell^+}+\cos\theta_{K\ell^-}$). The best candidate is selected by $\Delta E$.

The signal yield is extracted from a fit to the $M_{\rm{bc}}$ distributions. We have found no significant excess, and set the upper limits as follows:
\begin{eqnarray}
	{\cal B}( B \to X_s e^+ e^- ) &<& 10.2 \times 10^{-6} \ \ \ \ ( \rm{90\%~C.L.} ),\\
	{\cal B}( B \to X_s \mu^+ \mu^-) &<& 19.9 \times 10^{-6} \ \ \ \ ( \rm{90\%~C.L.} ).
\end{eqnarray}
These results are close to the SM predictions.~\cite{Inclusive}

\section{Summary}
We have studied penguin-mediated $B$ decays. The branching fractions of $B \to K^{*} \gamma$, $B^0 \to K_2^*(1430)^0 \gamma$, $B^+ \to K^+ \pi^+ \pi^- \gamma$, $B^+ \to K^{*0} \pi^+ \gamma$ were measured and the upper limit of $B^0 \to K^+ \rho^0 \gamma$ was set. The $B^+ \to K^+ \pi^+ \pi^- \gamma$ was observed for the first time. In $B \to K^{*} \gamma$ decay, we set the most stringent limit on the partial rate asymmetry, which was found to be consistent with zero. 

We observed $B \to K \ell^+ \ell^-$, and the measured branching fraction was found to be consistent with the SM predictions.~\cite{Exclusive} For other $b \to s \ell^+ \ell^-$ transitions, we set the upper limits of the branching fractions. These results were used to constrain the Wilson coefficients $C_9^{\rm{eff}}$ and $C_{10}$.~\cite{C9C10}

\section*{Acknowledgments}
We wish to thank the KEKB accelerator group for excellent operation of the KEKB accelerator.


\section*{References}
\begin{thebibliography}{99}
\bibitem{Belle} K.~Abe {\it et al.} (The Belle Collaboration), \Journal{\NIMA}{479}{117}{2002}.
\bibitem{KEKB} KEKB $B$ Factory Design Report, {\it{KEK Report}} 95-7 (1995), unpublished; Y.~Funakoshi {\it et al.}, Proc. 2000 European Particle Accelerator Conference, Vienna (2000).
\bibitem{CLEOAcpK2stgamma} T.~E.~Coan {\it et al.} (CLEO Collaboration), \Journal{\PRL}{84}{5283}{2000}.\label{ref:CLEOAcpK2stgamma}
\bibitem{BabarAcp} B.~Aubert {\it et al.} (The Babar Collaboration) \Journal{\PRL}{88}{101805}{2002}.\label{ref:BabarAcp}
\bibitem{K2stgamma} S.~Veseli and M.~G.~Olsson, \Journal{\PLB}{367}{309}{1996}; D.~Ebert, R.~N.~Faustov, V.~O.~Galkin and H.~Toki, \Journal{\PRD}{64}{3054001}{2001}; A.~S.~Safir, \Journal{\EPJ}{C15}{1}{2001}.
\bibitem{Xsgamma}K.~Abe {\it et al.} (The Belle Collaboration), \Journal{\PLB}{511}{151}{2001}.
\bibitem{Kll}K.~Abe {\it et al.} (The Belle Collaboration), \Journal{\PRL}{88}{021801}{2002}.
\bibitem{Exclusive} C.~Greub, A.~Ioannissian and D.~Wyler, \Journal{\PLB}{346}{149}{1995}; D.~Melikhov and N.~Nikitin, \Journal{\PLB}{410}{290}{1997}; A.~Ali, P.~Ball, L.~T.~Handoko and G.~Hiller, \Journal{\PRL}{D61}{074024}{2000}.
\bibitem{Inclusive} F.~Kr\"uger and L.~M.~Sehgal, \Journal{\PLB}{380}{199}{1996}; A.~Ali, G.~Hiller, L.~T.~Handoko and T.~Morozumi, \Journal{\PRD}{55}{4105}{1997}.
\bibitem{C9C10} A.~Ali, E.~ Lunghi, C.~Greub and G.~Hiller, arXiv:hep-ph/0112300.
\end{thebibliography}
\end{document}